\documentclass{aastex63}

\submitjournal{ApJL}

\shorttitle{Localized FRBs and Magnetars}
\shortauthors{Bochenek et al.}
\graphicspath{{./}{figures/}}
\usepackage{comment}
\begin{document}

\title{Localized FRBs are Consistent with Magnetar Progenitors Formed in Core-Collapse Supernovae}

\correspondingauthor{Christopher D.~Bochenek}
\email{cbochenek@astro.caltech.edu}

\author[0000-0003-3875-9568]{Christopher D.~Bochenek}
\affiliation{Cahill Center for Astronomy and Astrophysics, California Institute of Technology \\
1200 E.~California Blvd., MC 249-17 \\
Pasadena, CA 91125, USA}
%\affiliation{Owens Valley Radio Observatory, California Institute of Technology \\
%1200 E.~California Blvd., MC 249-17 \\
%Pasadena, CA 91125, USA}

\author{Vikram Ravi}
\affiliation{Cahill Center for Astronomy and Astrophysics, California Institute of Technology \\
1200 E.~California Blvd., MC 249-17 \\
Pasadena, CA 91125, USA}

\author{Dillon Dong}
\affiliation{Cahill Center for Astronomy and Astrophysics, California Institute of Technology \\
1200 E.~California Blvd., MC 249-17 \\
Pasadena, CA 91125, USA}
%\affiliation{Owens Valley Radio Observatory, California %Institute of Technology \\
%1200 E.~California Blvd., MC 249-17 \\
%Pasadena, CA 91125, USA}



\begin{abstract}

With the localization of fast radio bursts (FRBs) to galaxies similar to the Milky Way and the detection of a bright radio burst from SGR J1935+2154 with energy comparable to extragalactic radio bursts, a magnetar origin for FRBs is evident. By studying the environments of FRBs, evidence for magnetar formation mechanisms not observed in the Milky Way may become apparent. In this paper, we use a sample of FRB host galaxies and a complete sample of core-collapse supernova (CCSN) hosts to determine whether FRB progenitors are consistent with a population of magnetars born in CCSNe. We also compare the FRB hosts to the hosts of hydrogen-poor superluminous supernovae (SLSNe-I) and long gamma-ray bursts (LGRBs) to determine whether the population of FRB hosts is compatible with a population of transients that may be connected to millisecond magnetars. After using a novel approach to scale the stellar masses and star-formation rates of each host galaxy to be statistically representative of $z=0$ galaxies, we find that the CCSN hosts and FRBs are consistent with arising from the same distribution. Furthermore, the FRB host distribution is inconsistent with the distribution of SLSNe-I and LGRB hosts. With the current sample of FRB host galaxies, our analysis shows that FRBs are consistent with a population of magnetars born through the collapse of giant, highly magnetic stars.

\end{abstract}



\section{Introduction} \label{sec:intro}

The energetics and durations of fast radio bursts (FRBs) imply highly magnetized, compact progenitors. This has led to to magnetars being a leading explanation for the origins of FRBs \citep{LuKumar2018,Beloborodov2017,Metzgeretal2019}. The detection of a 1.5\,MJy\,ms radio burst from the Galactic magnetar SGR J1935+2154 is evidence that the extragalactic fast radio bursts (FRBs) originate from magnetars \citep{Bocheneketal2020b,CHIME1935}. 
%There are several arguments for this association. The burst from SGR J1935+2154 could have been seen from extragalactic distances. FRBs are mostly found in galaxies with some degree of star-formation, at least as much as the Milky Way, lending further support to this hypothesis \citep{Heintzetal2020}. \citet{Bocheneketal2020a}  predicted the fluence of the burst assuming it comes from the same population as extragalactic FRBs. The volumetric rate of SGR J1935+2154-like bursts matches the volumetric rate of FRBs. Finally, bursts like that from SGR J1935+2154 can reproduce the observed all-sky rate of FRBs \citep{Margalitetal2020}). 
This discovery, for the first time, allows us to study magnetars that exist beyond the Milky Way's sphere of influence. It provides us access to magnetars in a wide variety of galactic environments, and opens up the possibility of finding evidence for multiple channels of magnetar formation. 

\begin{figure}
    \centering
    \includegraphics{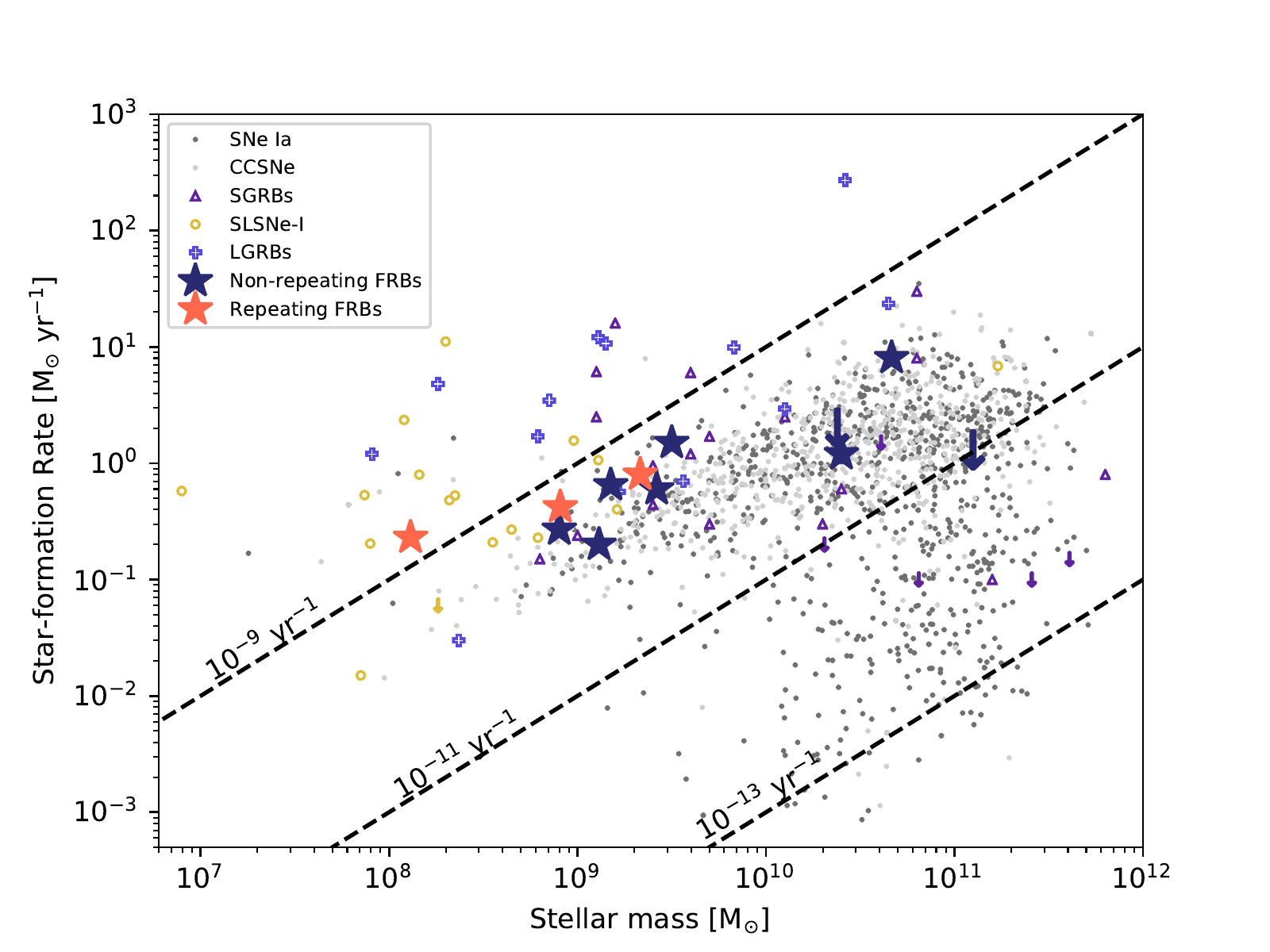}
    \caption{The star-formation rates and stellar masses of the hosts of a variety of different astrophysical transients that may correspond to evidence of different magnetar formation channels. The dark blue stars represent the nonrepeating FRBs while the orange stars represent the repeating FRBs. The light grey dots represent CCSNe in the GSWLC footprint with $z<0.3$, while the dark grey dots represent type Ia SNe with the same selection criteria. The CCSNe may track magnetars born through the fossil-field channel, while the Type Ia SNe may track millisecond magnetars born through AIC \citep{Margalitetal2019}. The open yellow circles are SLSNe-I \citep{Perleyetal2016}, while the open blue crosses represent long gamma-ray bursts (LGRBs) \citep{Levesqueetal2010}. These hosts may be representative of FRBs born as millisecond magnetars in engine-driven SNe. The short gamma-ray bursts are shown as open purple triangles \citep{Berger2014} and may track millisecond magnetars born through NS-NS mergers.}
    \label{fig:hosts}
\end{figure}

Several different mechanisms for magnetar formation have been proposed in the literature. These include magnetars born in normal core collapse supernovae (CCSNe) \citep{Schneideretal2019}, engine-driven CCSNe such as Type I superluminous supernovae (SLSNe-I) and long gamma-ray bursts (LGRBs) \citep{DuncanThompson1992,Woosley2010,KasenBildsten2010}, the accretion induced collapse (AIC) of a white dwarf \citep{DuncanThompson1992}, and the merger of two neutron stars (NS-NS) \citep{Giacomazzoetal2013,Giacamazzoetal2015}. Engine-driven SNe, AIC events, and NS-NS mergers are all different transient events expected to form proto-neutron stars with millisecond spin periods, which are hypothesized to create a convective dynamo that amplifies the magnetic field of the newly born neutron star to magnetar strengths. No such millisecond magnetar is required from the CCSNe formation channel (referred to as the ``fossil field'' channel), as the magnetar simply inherits its large magnetic field through conservation of magnetic flux from its progenitor star. Figure \ref{fig:hosts} shows the sample of FRB hosts in relation to different transient events that may track different magnetar formation channels.
%Magnetars can form in a number of different ways (AIC WD, WD+WD,NS+WD, NS+NS, CCSNe, SLSNe). 

The Milky Way population of magnetars is consistent with being dominated by normal CCSNe and there is not significant evidence for other formation channels in the Milky Way. Of the magnetars and magnetar candidates in the Milky Way \citep{OlausenKaspi2014,Espositoetal2020,Tendulkar2014}, 16 out of 30 are associated with a supernova remnant (SNR), massive star cluster, or star forming region. When restricted to the population of magnetars with a characteristic age $<10$\,kyr, a population for which a SNR is less likely to have dissipated and that has not had a significant amount of time to travel far away from its birth location, 12 out of 16 magnetars are associated with these objects. There is no indication that the Galactic magnetars originate from a special type of SN, favoring a ``fossil field" origin for their strong magnetic fields \citep{Schneideretal2019}. In an explosion driven by a magnetar engine, excess kinetic energy would be injected into the SNR from the spin-down of the magnetar, which would be observable at late times. However, the population of SNRs associated with magnetars does not appear different than the general population of SNRs in the Milky Way \citep{Zhouetal2019}. The millisecond magnetar hypothesis also predicts large kick velocities of order $10^{3}$\,km\,s$^{-1}$ \citep{DuncanThompson1992}. However, the distribution of magnetar kick velocities is similar to that of the general NS population \citep{Tendulkar2014,Delleretal2012,Dingetal2020}.  In addition, there is evidence that the stellar progenitors of magnetars span a wide range of masses \citep{Munoetal2006,Zhouetal2019}, while the progenitors of engine-driven explosions are likely more massive than a typical CCSN \citep{Blanchardetal2020}.

However, the most promising explanation for the rare SLSN-I explosions is an engine driven explosion powered by a millisecond magnetar \citep{Woosley2010,KasenBildsten2010}, as predicted by \citet{DuncanThompson1992}. Therefore, there is some evidence that there may be differences between the Galactic population of magnetars and the extragalactic population of magnetar. SLSNe-I are typically found in star forming regions of metal-poor dwarf galaxies, an environment that does not exist in the Milky Way \citep{Lunnanetal2015}. If a metal-poor environment with many highly massive stars is required to form millisecond magnetars in SLSNe-I, this \citep[together with the low SLSN-I rate;][]{Quimbyetal2013} would naturally explain them not being found in the Milky Way.
%would naturally explain the discrepancy with the Milky Way magnetar population. 

%It would be more difficult to reconcile a large discrepancy between the Galactic and extragalactic population of magnetars if there were a large number of extragalactic magnetars produced through either AIC or NS-NS mergers. 

Additionally, the Milky Way magnetar population does not appear to predominantly arise in AIC events or NS-NS mergers. These events track the stellar mass of a galaxy, while CCSNe track the star formation of a galaxy. The fraction of local-universe stellar mass contained in the Milky Way is approximately equal to the fraction of local-universe star formation occurring in the Milky Way \citep{Salimetal2007,KT2018,Bocheneketal2020a}. Therefore, if AIC or NS-NS mergers were as efficient at making magnetars as CCSNe, we would expect approximately half the magnetars in the Milky Way to be consistent with originating from AIC or NS-NS mergers. If this were true, due to the high kick velocities and long merger timescales of NS-NS mergers, we would expect to find a population of magnetars that is far away from the Galactic plane. If AIC is an efficient formation channel for magnetars, we would expect a large population of magnetars that are spatially uncorrelated with Galactic star-forming regions. Given that approximately 75\% of young Galactic magnetars are associated with star formation, we would expect magnetars born in AIC or NS-NS mergers to make up no more than 25\% of the extragalactic magnetar population. Furthermore, given that the volumetric rates of SLSNe-I \citep{Quimbyetal2013} and NS-NS mergers \citep{Abbottetal2017} are much lower than the magnetar birth rate \citep{KeaneKramer2008}, these channels cannot dominate the extragalactic magnetar population.

\citet{Giacomazzoetal2013} show that one possible outcome of a NS-NS merger is a stable, more massive NS. \citet{Giacamazzoetal2015} go on to show that the magnetic field of that stable NS can be amplified through a dynamo driven by internal turbulence. The signature of this population would be extragalactic magnetars with large offsets from their host galaxies and an association of extragalactic magnetars with massive quiescent galaxies. The host galaxies of these magnetars would have similar characteristics to the hosts of short gamma-ray bursts (SGRBs).

The accretion induced collapse of a white dwarf can produce the convective dynamo described in \citet{DuncanThompson1992}, amplifying the magnetic field of the newly born millisecond neutron star to magnetar strengths. AIC of a white dwarf can occur in several different ways, all of them involving an oxygen/neon white dwarf (ONe WD). An ONe WD can merge with another WD, or it can accrete material from a wide variety of companion stars, including AGB stars, helium giant stars, helium main sequence stars, red giant stars, or main sequence stars. \citet{Ruiteretal2019} explore the binary evolution of each of these pathways and predict delay time distributions for each of them. The signature of this formation channel would be a population of magnetars roughly consistent with the mass distribution of their host galaxies, and a large fraction would occur in quiescent galaxies. \citet{Margalitetal2019} approximate this population by assuming that the host galaxy properties are similar to that of Type Ia SNe. \citet{Margalitetal2019} also show that FRB 180924 \citep{Bannisteretal2019} is consistent with a magnetar born in a NS-NS merger or AIC event.

To date, there are 12 FRBs with secure host galaxy associations. In this paper, we use this sample to explore the possibility that there are multiple FRB progenitor channels. To test this possibility, we first need to check if FRB host galaxies are consistent with the expected hosts of the dominant magnetar formation channel in the Milky Way. One compelling way to understand this test, and analyses of FRB host galaxies more generally, is as a search for evidence of alternative magnetar formation channels. In Section 2, we describe our sample of FRB hosts, as well as the sample of CCSN, LGRB, and SLSN-I hosts we compare them to. We also describe a novel technique for comparing different samples of transient host galaxies that corrects for the fact that the underlying distribution of galaxies that different transients are sampled from is evolving with cosmic time. In Section 3, we will compare the population of FRB hosts to the transient hosts to test the hypothesis that other magnetar formation channels are needed to explain the hosts of FRBs. In Section 4, we discuss caveats to this work, and comparable studies, and conclude in Section 5. 

\begin{figure}
    \centering
    \includegraphics[width=0.49\textwidth]{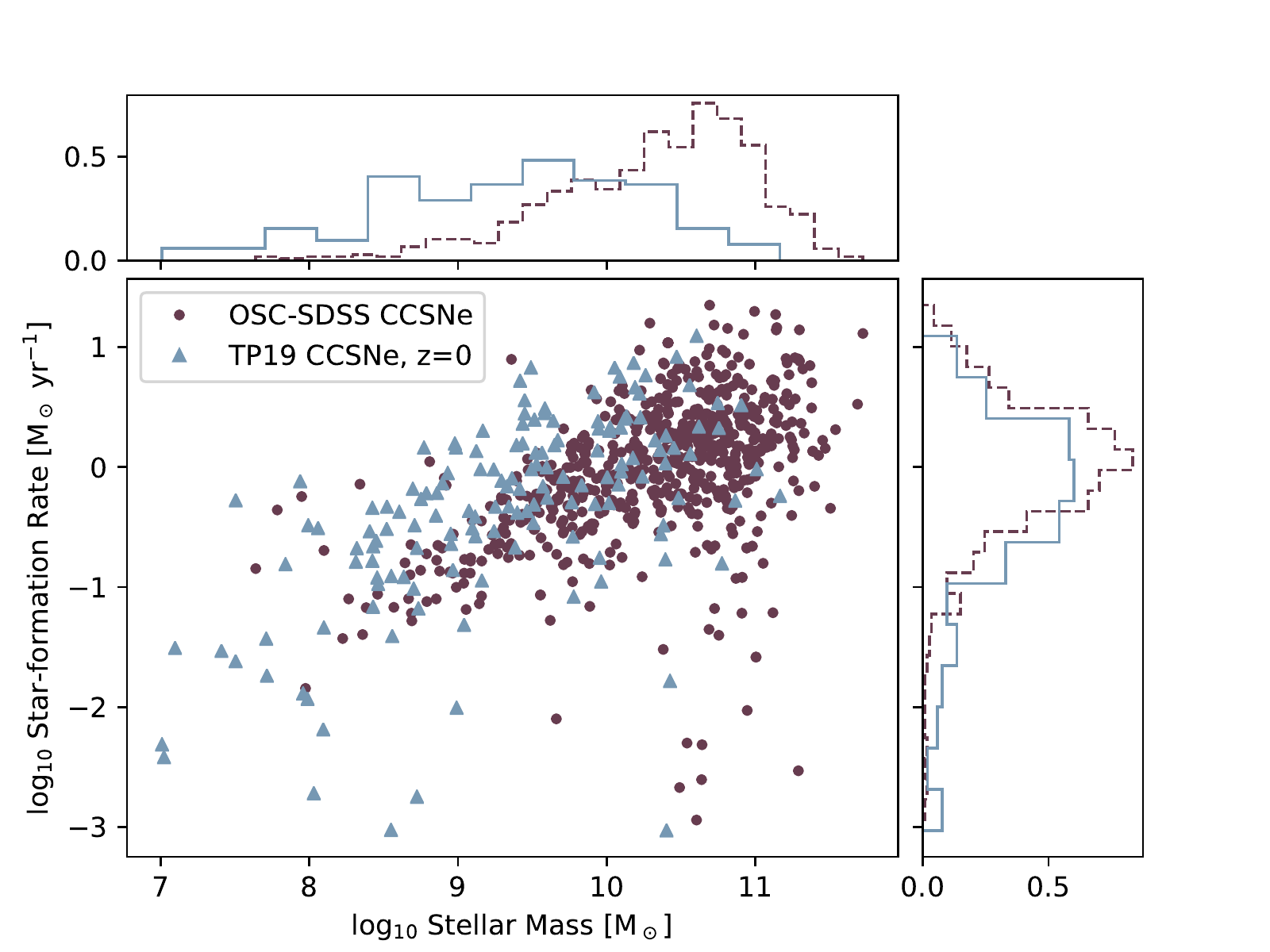}
    \includegraphics[width=0.49\textwidth]{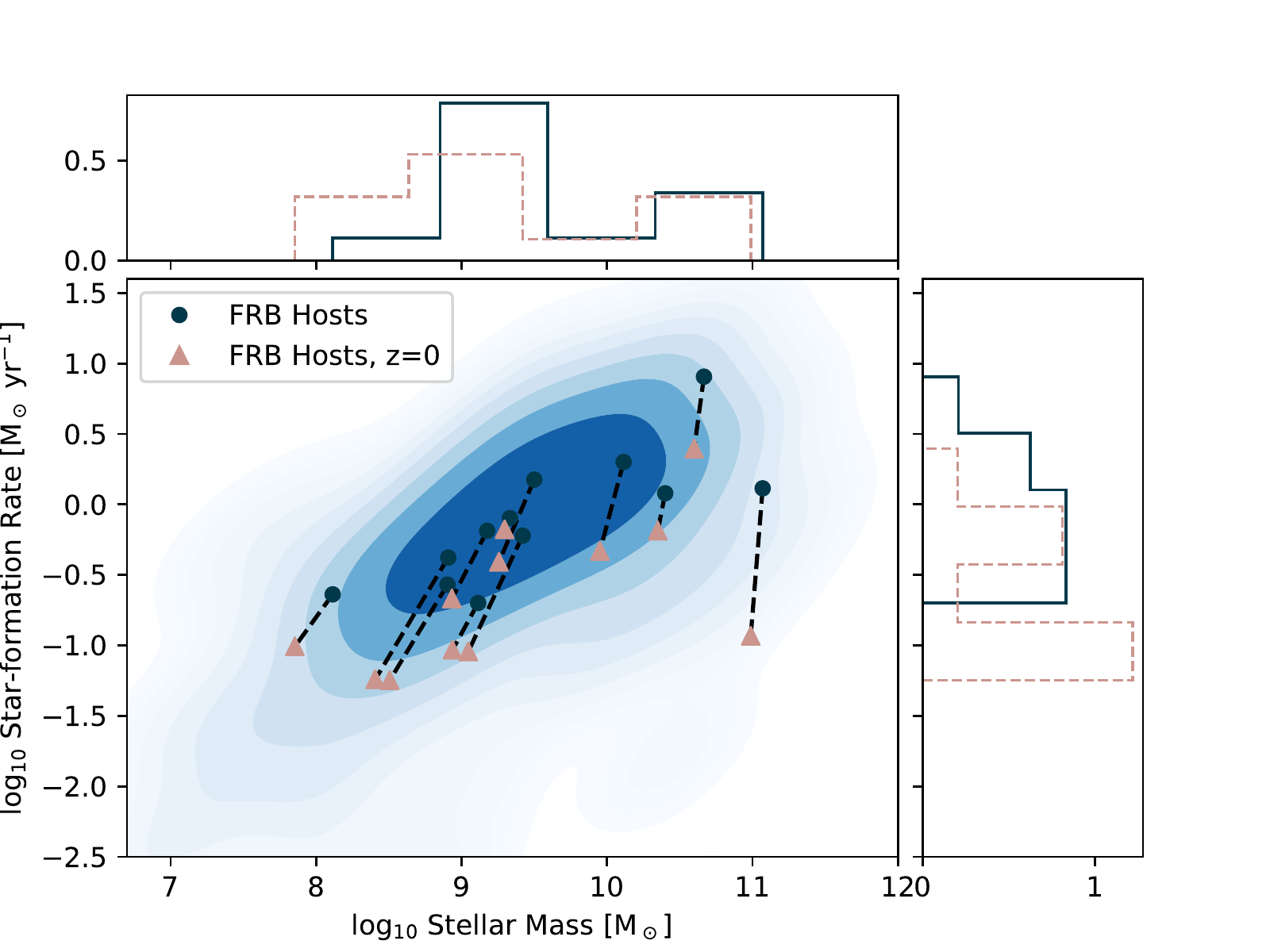}
    \caption{\textbf{Left}: The distribution of two different samples of CCSN hosts in SFR and stellar mass. The maroon dots correspond to CCSNe in the Open Supernova Catalog cross-matched with GSWLC galaxies, representing an incomplete sample of CCSN hosts. The blue triangles correspond to the hosts of the complete sample of CCSN hosts in TP19 that have been scaled to be representative of $z=0$ galaxies. \textbf{Right:} The dark blue dots correspond to the stellar masses and SFRs of FRB hosts without scaling them to be representative of $z=0$ galaxies, while the brown triangles correspond to the same FRB hosts scaled to be representative of $z=0$ galaxies. A dashed line connects each FRB in one sample to itself in the other sample. The blue contour plot is the kernel density estimate of the distribution of TP19 CCSN hosts, where the contours are logarithmically spaced across two orders of magnitude in probability.} 
    \label{fig:sample}
\end{figure}

\section{CCSN, SLSNe-I, LGRB, and FRB Host Samples}
\subsection{CCSN Host Sample}
We use the sample of CCSNe published in \citet{TP19}, henceforth referred to as TP19. This sample contains objects classified as Type II, Type IIP, Type IIb, Type Ib/Ic, Type IIn/Ibn, as well as two Type Ic-BL SNe. This sample of CCSNe is selected from the ASAS-SN supernova sample \citep{asassn}, an unbiased, magnitude limited survey. Given that ASAS-SN is a shallow survey, this sample has a median redshift of 0.014. The fact that this sample is at such low redshift means the host galaxy of every supernova has been identified, regardless of how small it is. 

The stellar masses and SFRs of this sample were derived from fitting the spectral energy distribution (SED) from the UV to NIR of each galaxy. If the SED had sufficiently blue colors and a spectrum was available, the luminosities of H$\alpha$ and [OII] were also included in the fit. The left panel of Figure \ref{fig:sample} shows the distribution in stellar mass and SFR of the CCSN in TP19 and all CCSNe in the Open Supernova Catalog \citep{Guillochonetal2017} with $z<0.3$ cross-matched with the GWSLC galaxy catalog \citep{Salimetal2016}, which covers 90\% of the Sloan Digital Sky Survey (SDSS) footprint. There is a clear bias towards large hosts in the OSC due to the incompleteness of SDSS at low stellar masses, demonstrating the need for the complete sample of TP19.

\subsection{SLSNe-I and LGRB host samples}
We use the sample of confirmed SLSNe-I and LGRBs from TP19. This sample is restricted to $z<0.3$ and contains only LGRBs with associated SNe Ic-BL and LGRBs with deep limits on a SN counterpart. The stellar masses and SFRs of this sample were derived from fitting the spectral energy distribution (SED) from the UV to NIR of each galaxy.

\subsection{FRB Host Sample}
We use the twelve published FRBs with host galaxies as our sample of FRB hosts \citep{Bassaetal2017,Bannisteretal2019,Prochaskaetal2019,Ravietal2019,Marcoteetal2020,Macquartetal2020,Heintzetal2020}. The redshifts of these FRBs range from $z=0.034$ to $z=0.66$. Each of these galaxies has a published stellar mass and star-formation rate (SFR), with the exception of FRB 180916.J0158+65 \citep{Marcoteetal2020}. Pending a more detailed analysis, we use the star-formation surface density reported in \citet{Marcoteetal2020} to integrate over the area of the galaxy and estimate a SFR of 0.8 M$_\odot$ yr$^{-1}$ for the host of FRB 180916.J0158+65.

Two of the FRB hosts only report upper limits on the SFR \citep{Bannisteretal2019,Ravietal2019} due to contamination of H$\alpha$ from an ionizing continuum that is harder than typical star-forming galaxies. These upper limits are consistent with the SFRs inferred from their H$\alpha$ luminosity and [OII] luminosity for FRB 180924 and FRB 190523, respectively. We choose to treat these upper limits on the SFR as the true value, as this contamination is unlikely to be recognized in the CCSN sample, given that most of the CCSN hosts do not have spectra (TP19)\footnote{We do not use the revised SFR for the FRB 190523 host from \citet{Heintzetal2020}. The photometry of the FRB 190523 host was fitted by \citet{Ravietal2019} to a stellar population synthesis model that yielded an SFR measurement consistent with the reported upper limit from spectroscopy. Additionally, \citet{Heintzetal2020} did not account for dust extinction of the H${\rm \beta}$ line.}.

We note that this sample includes three FRBs that are known to repeat \citep{Bassaetal2017,Marcoteetal2020,Kumaretal2020} and nine FRBs that were localized as one-off events and are not known to repeat \citep{Bannisteretal2019,Ravietal2019,Prochaskaetal2019,Macquartetal2020,Heintzetal2020}. It is possible that if repeating and nonrepeating FRBs are two distinct classes, the fact that two different localization strategies were used could introduce bias in the FRB host distribution. However, in the absence of evidence that these two FRB populations represent different progenitor classes \citep{Ravi2019}, we combine both the repeating FRBs and nonrepeating FRBs into one sample.

\subsection{Correcting for an evolving galaxy distribution}
When determining whether two populations of transient host galaxies are consistent with each other, it is important to keep in mind that the underlying distribution of galaxies and the location of star formation in the universe is evolving. A sample of star-forming galaxies at high redshift will preferentially have higher SFRs and masses than a sample of star-forming galaxies at $z=0$. When searching for evidence of alternate magnetar formation channels using events at a variety of redshifts, one is interested not in SFRs and stellar masses of FRB hosts themselves, but rather where they lie in relation to the distribution of star-formation at these different redshifts. We refer to the location of a sample of transient hosts in relation to the distribution of star formation as the sampling function. Without taking the evolution of the star-formation distribution into account, it is possible to mistake the bias towards higher SFRs at higher redshifts for an intrinsic difference in the hosts of FRBs and CCSNe.

For each sample of host galaxies, we correct for the evolution of star formation with cosmic time by scaling the masses and SFRs of each transient so that the statistical properties of each sample is representative of galaxies at $z=0$. To do this, we conserve the p-values in both stellar mass and SFR relative to the distribution of star-forming galaxies at the redshift of the transient and $z=0$. The distribution of star-formation as a function of stellar mass at a particular redshift is given in Equation \ref{eq:sfrp}, where $\frac{dn(z)}{dM}$ is the mass function of star-forming galaxies, $S_0(M_*, z)$ is the median SFR as a function of $M_*$ and $z$ (known as the star-forming main sequence), and $k$ is a normalization factor such that $\int p_z(M_*) dM_* = 1$. Given that star-forming galaxies are log-normally distributed about $S_0(M_*, z)$ and the scatter is constant with stellar mass \citep{Speagleetal2014}, this normalization also accounts for the difference between the median and mean of a log-normal distribution.
\begin{equation}
    p_z(M_*) = k \frac{dn(M_*,z)}{dM} S_0(M_*, z)
    \label{eq:sfrp}
\end{equation}
First, we need to account for the fact that at lower redshifts, sampling functions that simply track star formation are more likely at lower redshifts to produce transients in smaller galaxies than higher redshifts. We adjust the stellar mass of each galaxy so that $\int_{0}^{M_0}p_{z=0}(M_*) dM_* = \int_{0}^{M_{z_t}}p_{z=z_t}(M_*) dM_*$, where $p_z(M_*)$ is given by Equation \ref{eq:sfrp}, $M_0$ is the mass of that galaxy at $z=0$, $z_t$ is the redshift of the transient, and $M_z$ is the mass of that galaxy at the redshift of the transient. This adjusted mass represents the equivalent mass at $z=0$ that is selected by the sampling function at $z=z_t$. \citet{Moustakasetal2013} show there is no significant evolution in the number density of star-forming galaxies between $z=0$ and $z=0.65$, the redshift range of interest, so we ignore this redshift dependence. To find $\frac{dn(M_*,z)}{dM}$, we fit the Schechter function in Equation \ref{eq:schechter}, where $M_c$ is the cutoff stellar mass, $\phi_0$ is the normalization, and $\Gamma$ is the power-law index, to the number densities of star-forming galaxies between $z=0.2$ and $z=0.3$ reported in \citet{Moustakasetal2013}. 
\begin{equation}
    \log_{10}{\frac{dn(M_*,z)}{dM_*}} = \phi_0-M_c+\Gamma(M_*-M_c))-10^{M_*-M_c}\log_{10}(e)
    \label{eq:schechter}
\end{equation}
To perform the fitting, we drew $10^3$ realizations of the data assuming a split Gaussian uncertainty. We find the median parameters and $1\sigma$ distributions of the fit are $M_0 = 10.6 \pm 0.2$, $\phi_0 = 8.34_{-.05}^{+.03}$, and $\Gamma = -0.1_{-.25}^{+.09}$. For $S_0$, we used the ``preferred fit'' SFR-stellar mass relationship at redshift $z$ published in \citet{Speagleetal2014}. We compute $p_z(M_*)$ between $10^{6.5}$\,M$_\odot$ and $10^{12}$\,M$_\odot$.

Once we have determined the equivalent stellar mass of the galaxy at $z=0$, we then utilize the fact that star-forming galaxies are log-normally distributed about the star-forming main sequence and the scatter about the star-forming main sequence does not evolve with redshift \citep{Speagleetal2014}. To scale the SFRs to be representative of $z=0$ galaxies, we ensure that the host of the transient is the same number of standard deviations away from the star-forming main sequence at $z=z_t$ for $M_*=M_z$ as at $z=0$ for $M_*=M_0$. The right panel of Figure \ref{fig:sample} shows the corrections made to each FRB host galaxy. By applying both of these corrections, each transient is corrected for the effects of how the distribution of star-formation changes with cosmic time, allowing direct comparison between two samples of transient host galaxies at various redshifts.% and the sample of FRBs and CCSNe can be directly compared.

\section{Are FRB Hosts Similar to CCSN/LGRB/SLSN-I Hosts?}

\begin{table}[]
\centering
\begin{tabular}{l|llll}
            & M$_*$ p-value & SFR p-value & sSFR p-value & KDE p-value    \\ \hline
CCSN--FRB    & \textgreater{}.25 & \textbf{.035} & .14 & .546           \\
LGRB--FRB    & \textbf{.022} & \textgreater{}.25 & \textbf{.027} & .054 \\
SLSNe-I--FRB & \textbf{\textless{}0.001} & \textgreater{}.25 & \textbf{\textless{}0.001} & \textbf{\textless{}.0001}
\end{tabular}
\caption{This table shows the results of the various statistical tests performed to determine if two host galaxy samples are consistent with each other. We report Anderson-Darling p-values of the cumulative distributions in stellar mass, SFR, and sSFR between the samples and the p-value corresponding to the likelihood of randomly drawing a less probable sample than the FRB hosts from the kernel density estimate of the type of the hosts of the transient being compared to FRBs. Significant results are highlighted in bold.} 
\label{tab:1}
\end{table}

\begin{figure}
    \centering
    \includegraphics[width=0.99\textwidth]{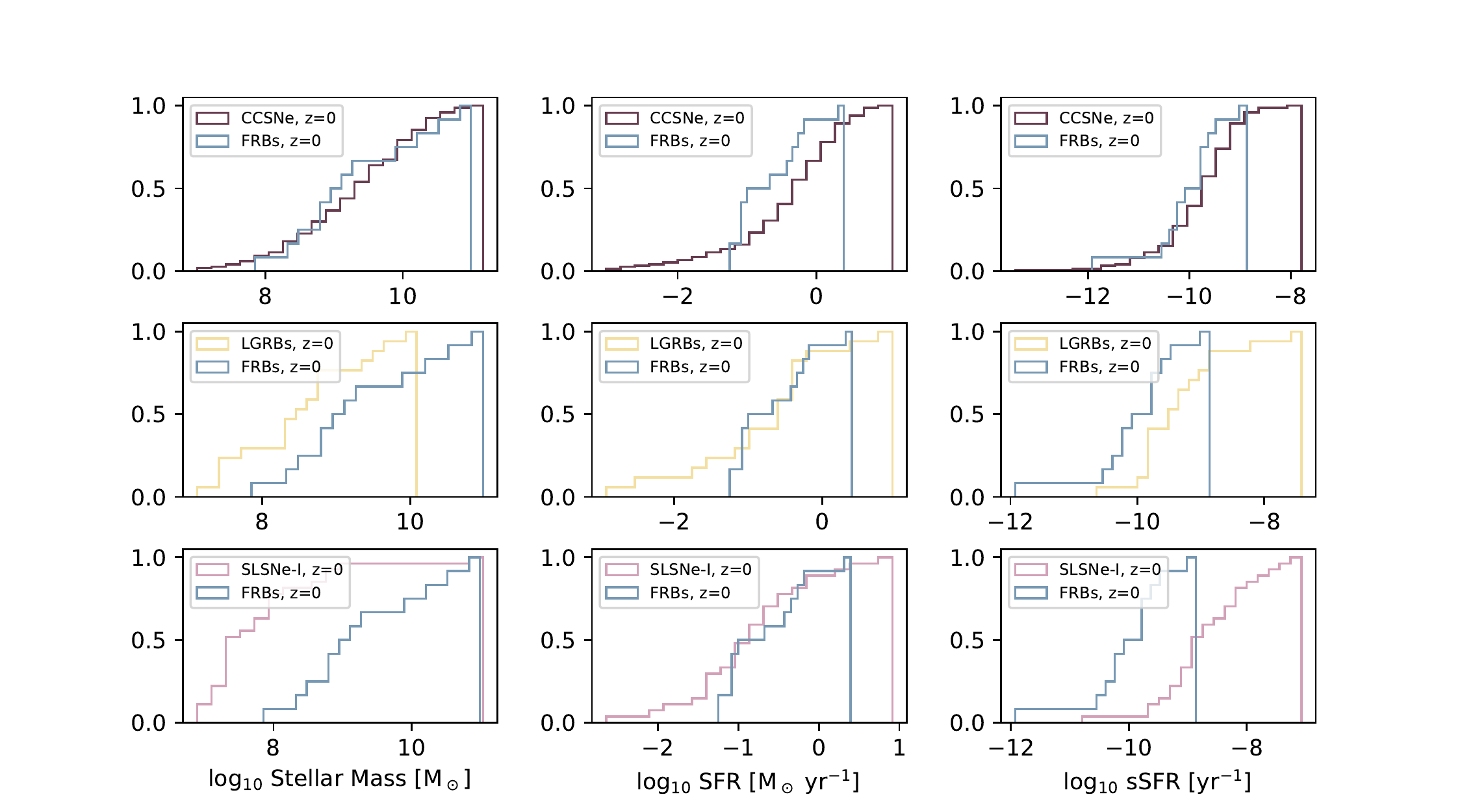}
    \caption{\textbf{Left:} The cumulative distributions of stellar masses of CCSNe, LGRB, SLSNe-I, and FRB hosts that have all been scaled to be representative of $z=0$ galaxies and FRB hosts that have been scaled to be representative of $z=0$ galaxies. \textbf{Middle:} The cumulative distributions of SFRs of CCSNe, LGRB, SLSNe-I, and FRB hosts that have all been scaled to be representative of $z=0$ galaxies and FRB hosts that have been scaled to be representative of $z=0$ galaxies. \textbf{Right:} The cumulative distributions of sSFRs of CCSNe, LGRB, SLSNe-I, and FRB hosts that have all been scaled to be representative of $z=0$ galaxies and FRB hosts that have been scaled to be representative of $z=0$ galaxies.}
    \label{fig:dists}
\end{figure}

After we scale the distributions of FRB, CCSN, LGRB, and SLSN-I hosts to be statistically representative of $z=0$ galaxies, we compare each cumulative distribution in stellar mass, SFR, and specific star-formation rate (sSFR) to the FRB hosts. These distributions are shown in Figure \ref{fig:dists}. We use the k-sample Anderson-Darling test \citep{AndersonDarling} to compare these distributions. Table \ref{tab:1} shows the results of each test. While the FRB host and CCSN host stellar mass and sSFR distributions are not significantly different, the SFRs of CCSN appear systematically higher. The FRB host distribution is inconsistent with the SLSNe-I host distribution with high confidence. Furthermore, the FRB host distribution has significantly higher masses and lower sSFRs that the LGRB host distribution.  %We use the k-sample Anderson-Darling test (CITATION) to compare these distributions between FRBs and CCSNe and find p-values of $0.19$, $>0.25$, and $0.07$ for their distributions in stellar mass, SFR, and sSFR, respectively.

We use a kernel density estimator with Gaussian kernels of widths determined by Scott's rule \citep{ScottsRule} on the logarithmic distribution of stellar masses and SFRs of the CCSN sample to estimate the distribution of these hosts in the stellar mass-SFR plane. We then compute the likelihood of the FRB sample from this distribution and compare this to the likelihood of $10^4$ random samples from this distribution. The random samples of this distribution resulted in a lower likelihood than the FRB hosts 54.6\% of the time. Therefore, even though the SFRs of FRB hosts and CCSN hosts are significantly different (Figure \ref{fig:dists}; Table \ref{tab:1}), we contend that FRB hosts and CCSN hosts are nonetheless consistent with being drawn from the same distribution. As demonstrated above, the sSFRs of FRB hosts and CCSN hosts are consistent, and the two-dimensional test better captures the available information. We repeat this procedure, comparing the LGRB hosts and SLSNe-I hosts to the FRB sample. A sample of galaxies drawn from the LGRB and SLSNe-I distributions have likelihoods smaller than the likelihood of the FRB host distribution 5.4\% and \textless{}0.01\% of the time, respectively. These results are also summarized in Table \ref{tab:1} Therefore, the hosts of FRBs are consistent with the CCSN host sample and marginally consistent with the LGRB hosts, but inconsistent with the SLSNe-I hosts. However, given that both the LGRB host stellar masses and SFRs are inconsistent with the FRBs, we disfavor a connection between FRBs and LGRBs.

%Given the observed FRBs \& magnetar formation channels, what can we say about the prevalence of magnetars formed in ways other than the Milky Way? If so, how common is this other channel?

%Most promising candidates for non-CCSNe: FRB 121102 (SLSNe), FRB 190711 (SLSNe), FRB 190611 (if no SFR in tiny host.)

%Are there any systematic biases (for example, high DMs in starbursts) that would bias our preference for one channel over another?

\section{Implications for FRB Progenitors and Magnetar Formation}

We have compared the host galaxies of FRBs to those of CCSNe, LGRBs, and SLSNe-I. Specifically, we considered the distributions of stellar masses and SFRs relative to the star-forming main sequence.   
%We have compared the distributions of stellar masses and SFRs relative to the star-forming main sequence of FRB hosts to LGRB hosts, SLSN hosts, and a complete sample of CCSN hosts. 
%This comparison was motivated by the possibility of multiple FRB progenitors and the possibility that magnetar formation channels other than the fossil field scenario happen.
We were motivated by the possibility of directly determining the origin and formation channels of FRB progenitors.
We find that the host galaxies of FRBs are consistent with the host galaxies of CCSNe, but not with the hosts of LGRBs and SLSNe-I.

The short durations and energetics of FRBs imply that they must originate from highly magnetized compact objects. The hosts of FRBs span a wide variety of galaxies, from dwarf galaxies with high sSFRs \citep{Bassaetal2017}, to massive starbursts \citep{Heintzetal2020}, to massive galaxies with SFRs below the star-forming main sequence \citep{Ravietal2019}. A Galactic magnetar has also produced an FRB \citep{Bocheneketal2020b,CHIME1935}.

All of these facts motivate the hypothesis that FRBs originate from magnetars like those found in the Milky Way.  This origin is consistent with our results that the hosts of FRBs are consistent with being drawn from the same selection function as CCSNe, and inconsistent with the SLSN-I and LGRB hosts. Magnetars born in CCSNe that do not have central engines dominate the population of Milky Way magnetars. Therefore, if magnetars are the dominant FRB progenitor, and the Milky Way magnetar population is representative, their host galaxies should be consistent with the hosts of CCSNe.

However, we caution that there may be systematic biases in the FRB population. Massive galaxies with high SFRs often have all of their star formation concentrated in a small region with an incredibly dense interstellar medium. For example, M82 is a nearby massive starburst galaxy which has a central starburst  of diameter 700\,pc that contains nearly all the star formation, and thus magnetars formed through corresponding channels. Using the H$53{\rm \alpha}$ flux in this region \citep{Puxleyetal1989}, the volume-averaged electron density of this region is $\sim 30$\,cm$^{-3}$. An FRB from M82 could have a dispersion measure of $10^4$\,pc\,cm$^{-3}$, likely also implying scatter-broadening orders of magnitude larger than $1$\,ms, making an M82 FRB impossible to detect in most FRB surveys. Indeed, it may not be a coincidence that FRB 191001, which is in a galaxy of similar mass and SFR as M82, is located on the outskirts of its host \citep{Heintzetal2020}. This bias away from galaxies with high SFRs may help explain the apparent discrepancy in SFR, but consistency in sSFR, between the CCSN hosts and FRB hosts.

With a sample of only twelve FRBs, we are only sensitive to dominant populations. It is possible that magnetars born in AIC events are a sub-dominant contributor to the population. Evidence for this channel would be an overabundance of FRBs in massive, quiescent galaxies. Furthermore, it is possible that one magnetar formation channel is significantly more likely to produce an FRB-emitting magnetar, as magnetars whose magnetic fields are formed in different ways may also dissipate that magnetic energy in different ways. Indeed, the host of FRB 121102 is very similar to the hosts of SLSNe-I \citep{Lietal2020}. Evidence for this hypothesis would be that repeating FRBs prefer galaxies with stellar masses $<10^9$ M$_\odot$ with high sSFRs, and low metallicities. The data would prefer a second FRB progenitor if, for example, a larger sample of FRB hosts requires that most magnetars are born through AIC events, the birth-rate of magnetars born in engine-driven SNe is much larger than the SLSNe-I/LGRB volumetric rate, or a significant fraction of FRBs are offset from their hosts.

In this paper, we have also demonstrated that it is necessary to use complete samples of transients to compare to FRB hosts due to significant biases induced by the completeness of galaxy catalogs. This often implies using only transients with relatively low redshifts. We have also developed a novel technique for comparing FRB host galaxy samples to samples of transients at lower redshifts by correcting for the evolution of how star formation is distributed with cosmic time. The advantages of this approach are that it makes it possible to directly compare host galaxy samples that have different redshift distributions to ensure that any differences between populations are representative of the nature of the transient event, rather than galaxy evolution. However, more work is required to compare the FRB host galaxy sample to transient host samples that do not track star formation, such as Type Ia SNe and NS-NS mergers. To do this analysis, it is necessary to incorporate the evolution of the quiescent galaxy population as well. Furthermore, this approach does not allow for determination the FRB delay-time distribution, another crucial piece of the FRB progenitor puzzle.

In a similar study, \citet{safarzadeh20} compare the host-galaxy stellar masses, SFRs, and projected offset distributions of the \citet{Heintzetal2020} FRBs with a population synthesis model for magnetars. The model links all magnetars to ongoing star-formation, consistent with a CCSN progenitor channel. \citet{safarzadeh20} conclude that the FRB hosts do not track star-formation activity, and hence are inconsistent with a magnetar origin, although this conclusion is not supported by the offset distribution. 
Although our conclusions are similar regarding star-formation alone, we come to a different conclusion regarding the overall consistency between FRB hosts and CCSN hosts. We argue that our results are more robust because: (\textit{i}) we directly compare FRB hosts with a complete, observational sample of CCSN hosts, thus obviating the need to rely on untested population synthesis models; and (\textit{ii}) we account for the two-dimensional distribution of host galaxies in star formation and stellar mass when reaching our final conclusion.

\section{Conclusions}

Using a novel approach for comparing two samples of transient host galaxies, we have determined that the current sample of FRB hosts is consistent with the hosts of CCSNe, and inconsistent with the hosts of SLSNe-I and LGRBs. This result is expected if magnetars similar to those in the Milky Way are responsible for the bulk of the FRB population. However, this result is limited by the current sample of FRB host galaxies. A larger sample of FRB hosts may turn up evidence for alternate magnetar formation channels or necessitate second progenitor for FRBs.

%Figures:
%1) Left panel: What types of galaxies is PAN-STARRS sensitive to, 1 hr on 5m class telescope, 1 hr on 10m class telescope at z~0.5. Right panel: What types of SFRs are GALEX/WISE sensitive to, 1 hr Halpha on 5m class telescope, 1 hr Halpha on 10m class telescope at z~0.5.

%Why would we not observe different progenitor channels in the Milky Way?

%What predictions do we make for a large sample of FRB host galaxies?

%How could further observations improve in usefulness to these types of studies? How should we be following up FRBs \& dealing with hostless ones?
%-GALEX band UV traces SFR out to 100 Myr, important for distinguishing intermediate and late delay times. NUV (230 nm) redshifts into SDSS u' band at z=0.53.
%-24/70 micron IR gets you out to 100 Myr
%-1.4 GHz/soft x-rays average emission is 100 Myr. How sensitive are these?

\acknowledgements

This research was supported by the National Science Foundation under grant AST-1836018. \software{Astropy \citep{Astropy2018}}.

\bibliography{sample63}{}
\bibliographystyle{aasjournal}



\end{document}